\documentclass{iopart}             
\usepackage[T1]{fontenc}
\usepackage[applemac]{inputenx}
\usepackage{graphicx}
\usepackage{url,graphics,epsfig}
\usepackage{amssymb}

\begin{document}
\setlength{\arraycolsep}{2.5pt}             
\jl{2}
%
%
%
\def\etal{{\it et al~}}
\def\newblock{\hskip .11em plus ....33em minus .07em}
%
%
%
%
%
\setlength{\arraycolsep}{2.5pt}             

\title[{\it K}-shell photoionization of Be-like Boron Ions]{{\it K}-shell photoionization of Be-like Boron (B$^+$) Ions: Experiment and Theory}

\author{A M\"{u}ller$^1\footnote[1]{Corresponding author, E-mail: Alfred.Mueller@iamp.physik.uni-giessen.de}$,
              S Schippers$^1$, R A Phaneuf$^2$, S W J Scully$^{2}$,
              A Aguilar $^{2,4}$, C Cisneros$^3$,
              M F Gharaibeh$^{2}\footnote[2]{Department of Physics, Jordan University of Science and Technology, Irbid, 22110, Jordan}$,
              A S Schlachter$^4$,
              and B M McLaughlin$^{5,6}\footnote[3]{Corresponding author, E-mail: b.mclaughlin@qub.ac.uk}$}

\address{$^{1}$Institut f\"{u}r Atom- ~und Molek\"{u}lphysik, Justus-Liebig-Universit\"{a}t Giessen,
                          Giessen, Germany}

\address{$^{2}$Department of Physics, University of Nevada,
                          Reno, NV 89557, USA}

\address{$^{3}$Centro de Ciencias F\'isicas, Universidad Nacional Aut\'onoma de M\'exico,
                           Apartado Postal 6-96, Cuernavaca 62131, Mexico}

\address{$^{4}$Advanced Light Source, Lawrence Berkeley National Laboratory,
                          Berkeley, California 94720, USA }

\address{$^{5}$Centre for Theoretical Atomic, Molecular and Optical Physics (CTAMOP),
                          School of Mathematics and Physics, The David Bates Building, 7 College Park,
                          Queen's University Belfast, Belfast BT7 1NN, UK}

\address{$^{6}$Institute for Theoretical Atomic and Molecular Physics,
                          Harvard Smithsonian Center for Astrophysics, MS-14,
                          Cambridge, MA 02138, USA}

%
%
\begin{abstract}
Absolute cross sections for the {\it K}-shell photoionization of Be-like boron ions were measured with
the ion-photon merged-beams technique at the
Advanced Light Source synchrotron radiation facility.
High-resolution spectroscopy with E/$\Delta$E up to 8800  ($\Delta$E $\sim$ 22 meV) covered the
energy ranges 193.7 -- 194.7 eV and 209 -- 215 eV.
Lifetimes of the strongest resonances are determined with relative uncertainties down to
approximately 4 \% for the broadest resonance. The measured resonance strengths are consistent
with 60~\% $\rm 1s^22s^2~^1S$ ground-state and 40~\% $\rm 1s^22s2p~^3P^o$ metastable-state ions in the primary ion beam
and confirmed by comparison with independent absolute photo-recombination heavy-ion storage-ring measurements
with B$^{2+}$ ions using the principle of detailed balance. Experimental determination of the line width
for the $\rm 1s2s^22p$ ~ $^1$P$^{\rm o}$  resonance gives a value of 47~$\pm$~2~meV and compares
favourably to a theoretical estimate of 47 ~meV from the R-matrix with pseudo-states (RMPS) method.
The measured line widths of the $\rm 1s2s2p^2$~$^3$P, $^3$D resonances are 10.0~$\pm$~2~meV
and  32~$\pm$~3~meV,  respectively,  compared to RMPS theoretical estimates of 9~meV and 34~meV.
\end{abstract}
%
%

\vspace{1.0cm}
\begin{flushleft}
Short title: {\it K}-shell photoionization of Be-like boron (B$^{+}$) ions\\
\submitted{\jpb:~ \today}
\end{flushleft}
\maketitle

%
%
%

\section{Introduction}

Satellites such as {\it Chandra} and  {\it XMM-Newton}
are currently providing a wealth of x-ray spectra on many
astronomical objects, but a serious lack
of adequate atomic data, particularly in the {\it K}-shell energy range,
impedes the interpretation of these spectra.
Spectroscopy in the soft x-ray region (0.5--4.5~nm), including
{\it K}-shell transitions of singly and multiply charged ionic
forms of atomic elements such as Be, B, C, N, O, Ne, S and Si, as well as L-shell
transitions of Fe and Ni, provides a valuable probe of the extreme
environments in astrophysical sources such as active galactic nuclei (AGN's),
x-ray binary systems, and cataclysmic variables
\cite{McLaughlin2001,Kaastra2011a,Brickhouse2010,Kallman2010,McLaughlin2013}. For example,
{\it K}-shell photoabsorption cross sections for the carbon isonuclear sequence
have been used to model the Chandra X-ray absorption spectrum of the bright blazar Mkn 421  \cite{McLaughlin2010}.

An important parameter in analyzing spectra from hot-gas or plasma environments is the charge-state distribution
of the chemical elements present in the observed radiation source.
The relative abundances of ions in different charge states of the key elements listed above are determined primarily by
photoionization (PI) \cite{McLaughlin2001,McLaughlin2013,Stancil2010,Miyake2010,Stancil2014}
and electron-ion collision processes~\cite{Mueller2008} including electron-impact ionization (EII)
and electron-ion recombination. Experimentally, EII has been mainly studied in laboratory-scale
colliding-beams experiment~(\cite{Borovik2013} and references therein) but also
in heavy-ion storage-ring experiments~\cite{Linkemann1995,Schippers2012}.
Electron-ion recombination is extensively studied at heavy-ion storage rings~\cite{Schippers2012,Schippers2010a}.
PI of ions has primarily been investigated using the photon-ion
merged-beam technique \cite{Phaneuf1999,Kjeldsen2006a}. Substantial amounts of PI data
have been obtained by employing the dual-laser-plasma technique~\cite{Kennedy2003}. Additional access to PI spectroscopy
is provided by the application of ion-trapping techniques \cite{Bizau2011a,Simon2010a}.

Experimental and accompanying theoretical PI studies on {\it K}-shell excitation
have been carried out so far on He-like Li$^\mathrm{+}$ \cite{Scully2006,Scully2007};
Li-like B$^\mathrm{2+}$ \cite{Mueller2010}, C$^\mathrm{3+}$ \cite{Mueller2009},  N$^{4+}$ \cite{Soleil2013},
Be-like C$^\mathrm{2+}$ \cite{Scully2005} and N$^{3+}$ \cite{Soleil2013},
B-like C$^\mathrm{+}$  \cite{Schlachter2004} and N$^{2+}$ \cite{Soleil2014},
C-like N$^\mathrm{+}$ \cite{Soleil2011},
N-like O$^{+}$ \cite{Kawatsura2002} and
F-like Ne$^{+}$ \cite{Yamaoka2001}.
Similarly, photoabsorption by neutral nitrogen \cite{McLaughlin2011} and
by neutral oxygen \cite{Stolte2013} has been studied in the vicinity of the {\it K}-shell ionization threshold.

The goal of the present  experimental and theoretical investigation is to provide accurate
values for photoionization cross sections, resonance energies, and natural linewidths resulting
from the photoabsorption of EUV radiation near the {\it K}-edge of Be-like boron.
This new study completes the work started at the ALS on {\it K}-shell excitation of
Li-like B$^\mathrm{2+}$ \cite{Mueller2010} and complements our previous experimental and
theoretical study on PI focusing on the valence shell of B$^+$ ions \cite{Schippers2003}.
Beside a pure basic scientific interest in the Z-dependence of PI via photoabsorption near the
{\it K}-edge of isoelectronic atoms and ions, astrophysical observations of boron provide motivation
for investigating this relatively light low-Z element. The abundances of the light elements are
valuable probes of both the early universe and the nucleosynthetic history of the Milky Way.
For this purpose, interstellar boron has been in the focus of numerous astrophysical observations.
Many studies have aimed at the determination of both the $^{11}$B/$^{10}$B isotope ratio~\cite{Federman1995}
and the boron abundance relative to hydrogen~\cite{Federman1993} along the lines of sight towards prominent stars.
A comprehensive survey of boron abundances in interstellar clouds together with an extensive compilation
of the relevant literature has been published by Ritchey \etal~\cite{Ritchey2011}. Since most of the boron
within the interstellar medium is singly ionized  the 136.2 nm B$^{+}$ line is most suitable for such studies.
In hot main-sequence B-type stars the 206.58 nm B$^{2+}$ line has been observed instead.
An overview of such measurements has been published by Proffitt and Quigley~\cite{Proffitt2001}.
The analysis of the observations requires detailed knowledge of atomic data such as oscillator strengths
and exact wavelengths of all species that might be observed in the given wavelength range~\cite{Proffitt99}.
Assessment of line strengths is a very critical issue as was recently pointed out by Bernitt \etal~\cite{Bernitt2012}.
Reliable photoionization cross section data for the Be-like boron B$^\mathrm{+}$ ion in the vicinity of the
{\it K}-edge are important for the modelling of specific astrophysical plasmas such as the
hot (photoionized and collisionally ionized) gas surrounding B-type stars.

Excitation of a Be-like  B$^+$  {\it K}-shell electron to a bound upper level or the removal
of a {\it K}-shell electron from neutral boron produces autoionizing states which can decay by Auger or
Coster-Kronig transitions.  Previously, identification of {\it K}-Auger transitions from singly
and multiply charged ionic states of boron has been performed experimentally by using
electron spectroscopy in ion-atom collisions \cite{bruch77,rodbro77,rodbro79,Benis2004},
photon absorption and emission by laser-produced plasmas \cite{kennedy78,Costello1992},
electron-impact ionization \cite{Hofmann1990}, beam-foil spectroscopy \cite{Ryabtsev2005}, high-resolution spark
spectroscopy \cite{Kramida2008} and high-resolution PI spectroscopy employing synchrotron
radiation \cite{Mueller2010}. Theoretically, resonance energies and linewidths for Auger
transitions in the Be-like boron ion have been calculated using a variety of methods, such
as 1/Z perturbation theory \cite{safronova69,Safron80,Vainshtein1993},
 unrestricted Hartree-Fock (UHF) \cite{Luken1983},
configuration interaction for initial and final states \cite{Serrao1986}, single or multi-configuration
Dirac Fock (MCDF) \cite{Costello1992}, the Saddle-Point-Method (SPM) with R-matrix or
complex coordinate rotation methods \cite{chung83,chung89,chung90,Lin2001,Lin2002},
complex scaling, multi-reference configuration interaction (MR-CI)~\cite{Yeager2012}
the spin-dependent localized Hartree-Fock density-functional
approach \cite{Chu2007} and R-matrix theory~\cite{Petrini1981}.

With the present report on PI of B$^\mathrm{+}$ ions in an ALS experiment, the first measurements of
the PI cross sections for the Be-like boron ion are presented for the photon energy range in the vicinity of the {\it K}-edge.
Theoretical results for PI near the {\it K}-edge were previously derived from the independent particle
model \cite{rm1979,verner1993,verner1995}, but do not account for autoionizing {\it K}-shell excited states.
To benchmark theory and obtain suitable agreement with the high-resolution accurate experimental
PI measurements performed at third-generation synchrotron light source facilities,
state-of-the-art theoretical methods are required. Theoretical approaches need to employ
highly correlated wavefunctions, be capable of including relativistic effects
(when appropriate) as well as radiation and Auger-damping effects.
In addition, when metastable states are present in the parent experimental ion beam,
further theoretical calculations are necessary to determine their contribution,
as in the present study for {\it K}-shell photoionization of Be-like boron B$^\mathrm{+}$ ions.
For handling all these requirements the state-of-the-art R-matrix method \cite{rmat,codes}
was employed in the present study. The experimental techniques were the same as those used in the
most recent experimental work at the ALS in PI studies on
Be-like ions \cite{Mueller2010b}, multiply charged iron ions \cite{Gharaibeh2011a},
singly charged argon \cite{Phaneuf2011} and singly charged krypton \cite{Hino2012} ions.

From a general point of view, studies like these are important in order to provide
accurate results for absolute photoionization (PI) cross sections, resonance energies
and natural linewidths for benchmarking results that update existing literature
values \cite{rm1979,verner1993,verner1995,Kasstra1993} and as such should be
used in preference to those that are currently in use in the various astrophysical
modelling codes such as CLOUDY  \cite{Ferland1998,Ferland2003},
XSTAR \cite{Kallman2001} and AtomDB \cite{Foster2012}.

Promotion of  a {\it K}-shell electron in Be-like boron B$^\mathrm{+}$ ions  to an outer
np-valence shell (1s $\rightarrow$ np)  from the ground or metastable levels produces
states that can autoionize, forming a B$^\mathrm{2+}$ ion and an outgoing free electron.
The strongest excitation process of this kind  in the interaction of a photon with the
$\rm 1s^22s^2~^1S$ ground-state of the Be-like boron ion is the 1s $\rightarrow$ 2p
photo-excitation with subsequent Auger decay of the intermediate doubly excited state
$$
 h\nu + {\rm B^{+}(1s^22s^2~^1S)}  \rightarrow  {\rm B^{+} ~ (1s2s^2 \,2p ~ ^1P^o) }
 $$
 $$
 \swarrow \quad \searrow
 $$
 $$
 {\rm  B^{2+}~(1s^22s~^1S) + e^- ({\it k^2_{\ell_1}})} {\quad} {\rm or} {\quad} {\rm  B^{2+}~(1s^22p~^2P) + e^-({\it k^2_{\ell_2}})},
$$
where ${\it k^2_{\ell_i}}$  ($i=1,2$) represent the energies of the outgoing Auger electrons from the two different decay processes, respectively.
 B$^{+}$(1s2p$^3$ $^1$P$\rm ^o$)  and B$^{+}$(1s2s$^2$np $^1$P$\rm ^o$) states (n $\geq$ 3)
can also be excited with considerable probability and subsequently decay by electron emission.
In the case of the $\rm 1s^22s2p~^3P^o$ metastable state the dominant  autoionization
processes caused by the 1s $\rightarrow$ 2p photo-excitation process  are
$$
 h\nu + {\rm B^{+}(1s^22s2p~^3P^o)}  \rightarrow  {\rm  B^{+} (\{1s2s[^{1,3}S]\,2p^2~^3P,^1D,^1S\}~^{3}P, ^{3}D, ^{3}S)}
$$
$$
\downarrow
$$
$$
{\rm B^{2+} (1s^22s~ ^2S) + e^-.}
$$
From the metastable level of Be-like boron, the terms B$^{+}$(1s2s[$^{1,3}$S]2pnp\, $^3$D, $^3$P, $^3$S)
can also  be excited. The inner-shell autoionization resonances created (by the above processes)
appear in the corresponding PI cross sections (in the energy region near to the {\it K}-edge) on top of
a comparatively small continuous background cross section for direct photoionization of a 2s electron.
The present investigation provides absolute values (experimental and theoretical) for
PI cross sections, resonance energies and the natural  linewidths for a number of these terms.

The principle of detailed balance can be used to compare the present PI cross-section
measurements with experimental cross sections for the time-inverse photo-recombination (PR) process.
This comparison provides a valuable cross check between entirely different
experimental approaches used for obtaining atomic cross sections on absolute scales.

The layout of this paper is as follows. Section 2 presents a brief outline of the theoretical work.
Section 3 details the experimental procedures used. Section 4 presents a discussion of the
results obtained from both the experimental and theoretical methods. It includes a comparison
of the present PI work with an observation of electron-ion recombination of B$^{2+}$(1s$^2$2s~$^1$S)
at the heavy-ion storage ring TSR of the Max-Planck-Institute for Nuclear Physics in
Heidelberg, Germany~\cite{Boehm2006,Lestinsky2004}.
Finally in section 5 conclusions are drawn from the present investigation.

\section{Theory}\label{sec:theory}
\subsection{R-matrix}
In R-matrix theory, all photoionization/photoabsorption calculations require  the generation
of atomic orbitals based primarily on  the structure of the residual ion \cite{Burke2011,McLaughlin2010,rmat,codes,damp}.
PI cross-section  calculations for Be-like  boron B$^\mathrm{+}$ ions were performed in $LS$ coupling
within the confines of the R-matrix approach \cite{Burke2011,McLaughlin2010,McLaughlin2001,rmat,codes,damp}.
We included 249 levels of the B$^\mathrm{2+}$ residual ion in the close-coupling expansion
of the wave functions used in our work. For the case of ground state we investigated
the inclusion of a larger number of states (390 levels) in our CI model to check on the convergence of our results.
An n=4 basis set of B$^\mathrm{2+}$ orbitals was employed  which was
constructed using the atomic-structure code CIV3 \cite{Hibbert1975} to represent the wave functions.
Due to the presence of metastable states in the parent ion beam used in the experiment, PI cross-section
calculations were required to be performed for both the $\rm 1s^22s^2~^1S$ ground state and the $\rm 1s^22s2p~^3P^o$
metastable state of the B$^\mathrm{2+}$ ion.  All the PI cross section
calculations were carried out in $LS$ - coupling.

For the structure calculations of the residual B$^\mathrm{2+}$ ion,
all physical orbitals were included up to n=3 in the configuration-interaction
wave-function expansions used to describe the states. The Hartree-Fock $\rm 1s$ and $\rm 2s$ tabulated
orbitals of Clementi and Roetti \cite{Clementi1974} were used together with the 2p and n=3 orbitals
which were determined by energy optimization on the appropriate spectroscopic state using the
atomic structure code CIV3 \cite{Hibbert1975}.  The n=4 correlation (pseudo) orbitals were
determined by energy optimization on the ground-state of the B$^{2+}$ ion in order to account
for core relaxation and additional correlation effects in the multi-reference configuration interaction wave functions.
This corresponds to excitations of the ground state electronic configuration as well as from excited states.
The residual ion states were then represented by using these multi-reference configuration
interaction wave functions.

The experimental energies of the hole states (which the prominent B$^\mathrm{+}$ Rydberg resonances converge to)
are in suitable agreement with our current theoretical estimates.
The recommended value for the B$^\mathrm{2+}$ ($\rm 1s2s^2 ~^2S$) K-edge,
relevant to the ground state located at 192.731 eV \cite{NIST2013,Kramida2008},
compares favourably with our theoretical estimate of 192.528 eV (a difference of  0.1 \%).
Similarly for the excited  B$^\mathrm{2+}$ ($\rm 1s2s2p ~^4P^{o}$) hole state,
the recommended value of 194.75057 eV \cite{NIST2013,Kramida2008}
once again compares well with our theoretical estimate of 194.76067 eV.

The non-relativistic $R$-matrix approach was used to calculate the energies of the ground and metastable
B${^\mathrm{+}}$ ion states and the subsequent PI cross sections.  Since metastable states
are present in the parent ion beam in the experiment, PI cross sections
for the B$^\mathrm{+}$ ($\rm 1s^22s^2$\, $\rm ^1$S) ground-state and the
B$^\mathrm{+}$ (1s$\rm ^2$2s2p $\rm ^3$P$\rm ^o$) metastable state were
carried out for all total angular momentum scattering symmetries that contribute to the total
PI cross section under the usual dipole selection rules.

Two-electron promotions out of selected base configurations of the
B$^\mathrm{+}$ ion into the orbital set were allowed in order to generate correlated
multi-reference configuration interaction scattering wave-functions.
The cross section calculations were performed with twenty continuum functions and a
boundary radius of 13.2 Bohr radii to accommodate the diffuse n=4 pseudo-orbitals that were
included to account for core relaxation and electron correlation effects.

\subsection{Scattering}
An efficient parallel version of the R-matrix programs \cite{rmat,codes,damp,ballance06}
was used to determine the single photoionization cross sections.
For the $\rm ^1S$ ground state and the  $\rm ^3P^o$
metastable state the outer region electron-ion scattering problem
was solved (in the resonance region below and  between all the thresholds) using an extremely
fine energy mesh of 2$\times$10$^{-7}$ Rydbergs ($\approx$ 2.72 $\mu$eV) to fully resolve
all the fine resonance features in the {\it K}-shell PI cross sections.
Radiation and Auger damping were also included in our calculations.
The multi-channel R-matrix  eigenphase derivative (QB) technique
(applicable to atomic and molecular complexes) of Berrington and
co-workers \cite{keith1996,keith1998,keith1999} was used to determine the
resonance parameters. The resonance width $\Gamma$ was determined from
the inverse of the energy derivative of the eigenphase sum $\delta$ at the resonance energy $E_r$ via
\begin{equation}
\Gamma = 2\left[{\frac{d\delta}{dE}}\right]^{-1}_{E=E_r} = 2 [\delta^{\prime}]^{-1}_{E=E_r} \quad.
\end{equation}
Finally, in order to compare directly with experiment, the theoretical
cross sections were convoluted using a Gaussian profile function of an
appropriate full width at half maximum (FWHM). An admixture of  40 \% metastable
 and 60 \% ground state ions was used to simulate the experimental measurements  performed
at the ALS.

\section{Experiment}\label{sec:exp}

The ion-photon-beam (IPB) end-station of the ALS undulator beamline 10.0.1 
was used for the present experiment. The experimental arrangement at the IPB has been previously  described in detail
by Covington and co-workers \cite{Covington2002}. The experimental 
procedures used in the present study on Be-like B$^{+}$ ions were similar
to those utilized in our previous measurements on the {\it K}-shell PI cross sections
for multiply charged boron \cite{Mueller2010} and carbon ions \cite{Schlachter2004,Scully2005,Mueller2009}.

A compact all-permanent-magnet electron-cyclotron-resonance (ECR) ion 
source~\cite{Broetz2001} was used to generate the required B$^{+}$ ions. 
The working substance in the ion source was gaseous BF$_3$ 
introduced by a fine-regulation valve.  B$^{+}$ ion beams with electrical currents of
about 200~nA at energy 6~keV  were extracted from the ion source which resided on a positive potential of +6~kV.
By passing the ion beam through a dipole magnet the desired parent-ion 
charge-to-mass ratio was selected using a suitably positioned slit arrangement. 
The collimated ion beam was then merged with and centered onto the 
counter-propagating photon beam by applying a spherical electrostatic 
deflector, the merger,  and several electrostatic ion-beam steering and focussing devices. Downstream of  the
interaction region, the ion beam was bent off the photon-beam axis by a second
dipole magnet, the de-merger,  that also separated the ionized B$^{2+}$ product ions from the B$^{+}$ parent
ion beam. The B$^{2+}$ ions were counted with 97\% efficiency employing a well characterized single-particle detector,
and the B$^{+}$ ion current was collected by a large Faraday cup inside 
the de-merger and monitored with a sensitive amperemeter for normalization purposes. The measured B$^{2+}$
count rate was partly due to the PI events under study and partly resulted from collisions 
of the primary ions with residual gas particles or metal surfaces.  
Separation of the true photon-induced signal from the background events was 
accomplished by  mechanically chopping the photon beam and by 
chopping-phase sensitive recording of the B$^{2+}$ ion count rate.

For the measurement of absolute cross sections it was also necessary to 
monitor the photon flux and to quantify the photon-ion beam overlap. 
In a first step the beam overlap was optimized by using commercial rotating-wire beam-profile 
monitors and movable slit scanners. Once a sufficiently robust overlap was accomplished 
the same equipment was used to measure the formfactor~\cite{Phaneuf1999} 
which quantitatively describes the beam overlap.
The photon flux was measured with a calibrated photodiode. 
After subtracting the background from the measured B$^{2+}$ ion count 
rate the resulting photoionization signal was normalized to the measured ion
current, the photon flux and the form factor. 
Absolute cross sections and resonance strengths for photoionization have thus
been obtained with an estimated uncertainty of $\pm$20\% in the energy range 193.7--194.7~eV.
Since a considerable effort is required for carrying out reliable absolute cross-section measurements
these were only performed at the peaks of the dominant PI resonances.
The low counting rates obtained in the energy range 209.5--215.0~eV
did not allow us to carry out absolute cross-section measurements.
Data in that energy range were normalized to the present theory.

Detailed high-resolution energy-scan measurements were performed by stepping the photon energy
through a preset range of values. The desired resolving power was preselected by
adjusting monochromator settings of the beam-line  accordingly. 
The scan measurements provide relative photoion yields and
therefore had to be normalized to the absolute data points. 
The photon-energy scale was calibrated by carrying out photoabsorption measurements
with SF$_6$ and Ar gas for the well known resonance features \cite{Hudson1993,King1977} at
energies 176--182~eV and 242--251~eV, respectively. Monochromator settings resulting from the calibration points in
these ranges were linearly interpolated to obtain the scaling factors for measured
energies in the present range of interest. Since the parent ions 
counter-propagated the photon beam with a velocity of about 0.1\% of the speed of light, a Doppler correction
had to be carried out in order to obtain the nominal photon energy in 
the ions' rest frame. Subsequently the calibration factor was applied 
resulting in precisely determined photon energies with an estimated uncertainty of
at most $\pm 30$ meV in the range of the present measurements. 
The possible calibration error almost exclusively determines the 
absolute uncertainties of the resonance energies resulting from 
the present study. Relative peak positions are obtained with a statistical uncertainty of the order of only 1 meV.

%
%
%
\begin{figure}
\begin{center}
\includegraphics[width=\textwidth]{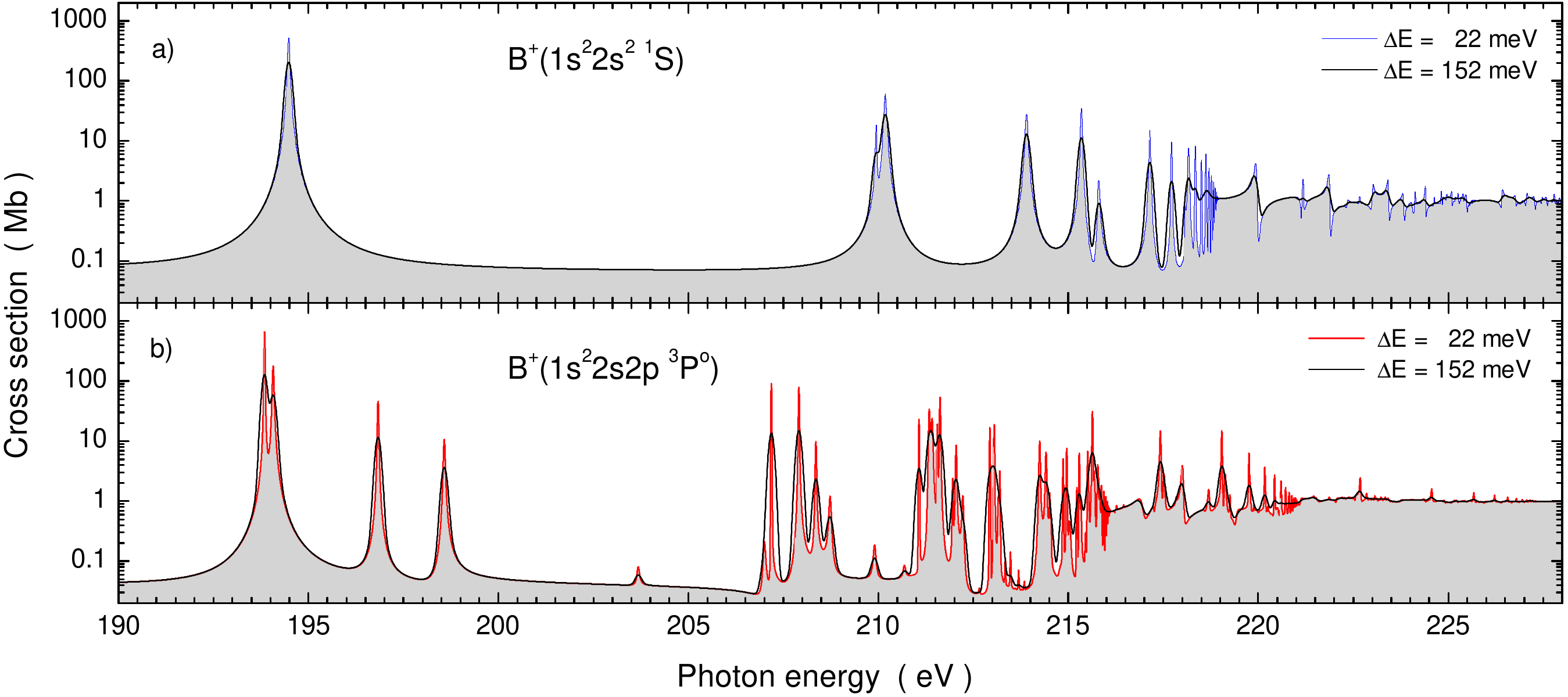}
\caption{\label{fig:theoverview}  Theoretical cross sections for {\it K}-shell photoionization (PI) of Be-like B$^\mathrm{+}$ ions
                                            for the photon energy  region 190--230 eV.
                                            The 249-state $LS$-coupling R-matrix calculations
                                            were carried out for a) ground-state and b) metastable-state parent ions and convoluted
                                            with Gaussian distributions of 22 meV and 152~meV at FWHM
                                            (the upper and lower limits of the resolution in the current ALS experiments).
                                            This procedure indicates the prominent resonance features that should be observed in the ALS experiments.}
\end{center}
\end{figure}

\begin{figure}
\begin{center}
\includegraphics[width=0.6\textwidth]{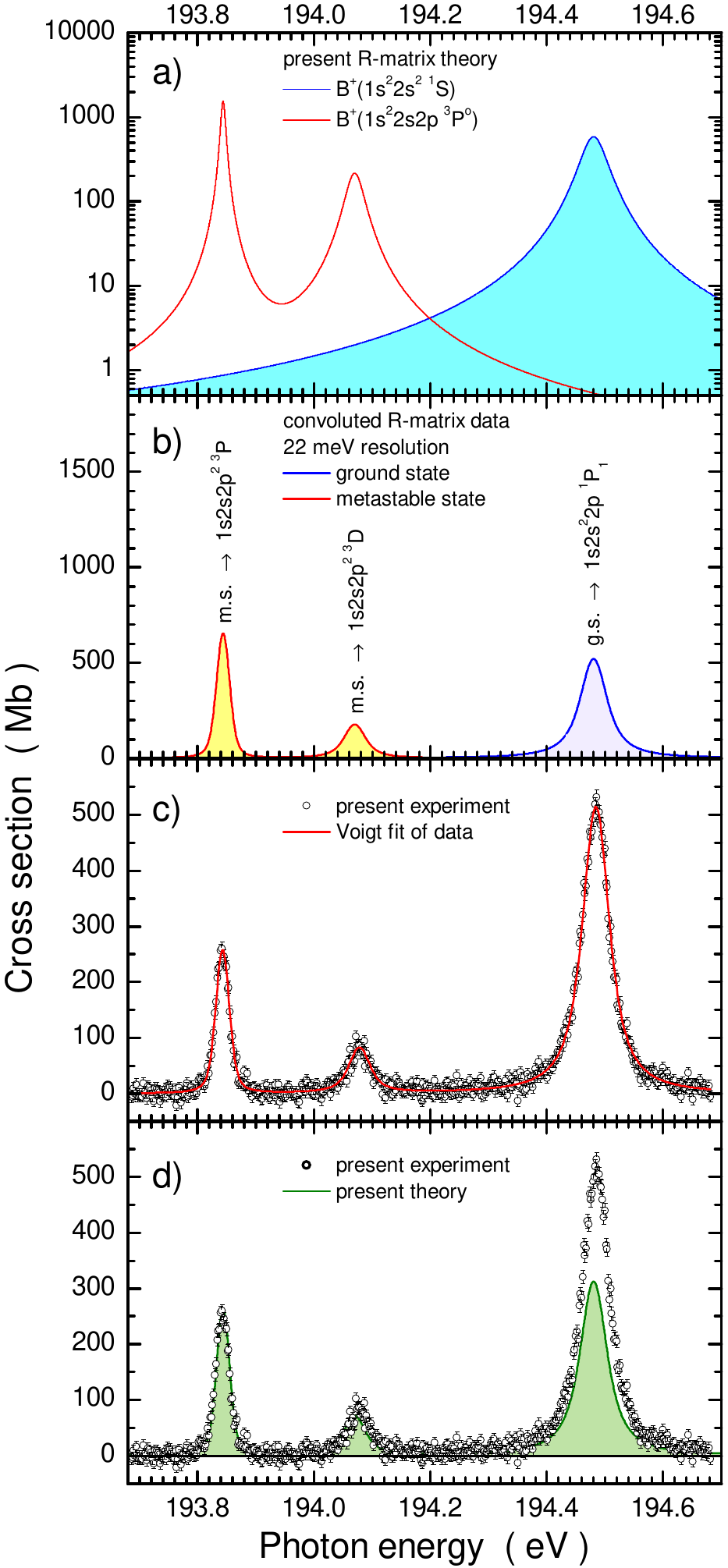}
\caption{\label{fig:Fig22meV}  Experimental and theoretical cross sections for {\it K}-shell
					photoionization (PI) of Be-like B$^\mathrm{+}$ ions.
                                            The present R-matrix results a) for (g.s.) ground-state (dark solid line with shading)
                                            and (m.s.) metastable-state contributions (light solid line) were convoluted with
                                            a 22~meV FWHM Gaussian distribution function  to simulate the ALS experimental
                                            resolution and are then shown on a linear scale in b) together with peak designations.
                                            The open circles in c) are the experimental results, the solid line is a Voigt fit to the data.
                                            A comparison of the experimental data and the R-matrix results
                                            weighted as 60\% ground and 40\% metastable states is provided in d).}
\end{center}
\end{figure}

\section{Results and Discussion}\label{sec:res}
\subsection{Photoionization}
Figure~\ref{fig:theoverview} provides an overview of the present R-matrix PI calculations carried
out for the B$^{+}$(1s$^2$2s$^2$~$^1$S) ground-state (upper panel) and the B$^{+}$(1s$^2$2s2p~$^3$P$^{\rm o}$)
metastable-state components (lower panel) of the parent ion beam in the present experiment.
The results in figure \ref{fig:theoverview} were convoluted at 22 meV
and 152 meV (the highest and lowest experimental resolution),
to highlight the prominent resonance features that can be observed in the cross sections.
The cross-section axis is on a logarithmic scale so that the many resonance features in the spectra can be distinguished.
Theoretical resonance parameters were inferred from the natural (unconvoluted) cross sections.
The results  of that analysis are given in tables 1 and 2. It is interesting to note that the
cross section for direct photoionization of the B$^+$ ion jumps by more than an order
of magnitude to about 1~Mb for both the ground state  and
the metastable state once direct {\it K}-shell ionization becomes possible.

In figure~\ref{fig:Fig22meV}, a comparison of the experimental and theoretical results is shown
for the photon energy range 193.60 eV to  approximately 194.47 eV. Panel a) displays the
natural cross sections from the R-matrix calculation on a logarithmic scale
which reveals the slight asymmetries resulting from interference effects. The results distinguish
between contributions from the ground state and the metastable state. Panel b) provides the
data from a) after convolution with a Gaussian distribution function of 22~meV full width at half maximum.
By this convolution experimental measurements as shown in panels c) and d) are simulated.
It is interesting to note that the data sets of panels a) and b) are very similar to one another
(not shown on identical scales) which indicates that an experiment at 22~meV resolution
is very sensitive to the natural line widths of the resonances in this energy region.
Indeed, the Voigt fit of the  experimental data at 22 meV resolution in c) provides the
line widths, and hence the lifetimes of the three resonances, with good accuracy
as shown in table~1. Panel d) provides a direct comparison of the present experiment
with convoluted R-matrix calculations assuming a 40\% fraction of $^3$P$^{\rm o}$
metastable B$^+$ ions admixed to 60\% of $^1$S ground-state ions in the parent ion
beam.
%
%
%

\begin{table}
\caption{\label{tab:fit1} Resonance energies $E_{\rm ph}^{\rm (res)}$ (in eV),
         natural linewidths $\Gamma$ (in meV) and resonance strengths $\overline{\sigma}^{\rm PI}$ (in Mb eV)
         for B$^{+}$ ions  in the energy range 193.60 eV to 194.7 eV. Although the present experimental
         uncertainty of energies on an absolute scale is $\pm 30$~meV,  three decimal figures are
         provided for the resonance energies  because the experiment is sensitive to the 1~meV level
         and allows one to determine energy spacings with uncertainties of the order of 2~meV.
         The present  experimental and theoretical investigations are compared with previous studies.
         The resonance strengths given are obtained for a mixture of 60\% ground-state
         and 40\% metastable ions in the parent ion beam. }
\begin{indented}
 \lineup
 \item[]\begin{tabular}{ccr@{\,}c@{\,}llcl}
\br
 Resonance    & & \multicolumn{3}{c}{ALS/Others}   & \multicolumn{1}{c}{R-matrix} & \multicolumn{2}{c}{
 Others}\\
 (Label)      & & \multicolumn{3}{c}{(Experiment)}      & \multicolumn{1}{c}{(Theory)} & \multicolumn{2}{c}{(Theory)}\\
 \ns
 \mr
 $\rm 1s2s2p^2$ ~ $^3$P						 & $E_{\rm ph}^{\rm (res)}$
               & 193.843   &$\pm$  & \00.03$^{\dagger}$  		& 193.844$^{a}$  	    & 	 	 &193.873$^{b}$ \\
 &           &193.86\phantom{0}  &$\pm$  &\00.20$^e$		        		&				    &		&			   \\
  \\
 & $\Gamma$
             & \phantom{0}10.0\0\0  	&$\pm$  & \02$^{\dagger}$     	  & \0\09.02$^{a}$ 		    & 	 	 &\0\09.82$^{b}$\\
            &                       &              &                                      &  &\0\09.36$^{g}$              &      	 & \0\09.74$^{h}$\\
            &                       &              &                                      & &                              	    &      	& \\
   \\
 & $\overline{\sigma}^{\rm PI}$
             &  8.85\phantom{0}    &$\pm$  & 1.8$^{\dagger}$      	& \0\08.44$^{a}$         	    &	  	 &		 \\
  \\
 \\
 $\rm 1s2s2p^2$ ~ $^3$D						 	& $E_{\rm ph}^{\rm (res)}$
              	& 194.078	&$\pm$  &\00.03$^{\dagger}$   	& 194.072$^{a}$    		   & 	  	& 194.125$^{b}$   \\
  &         	 &194.10\phantom{0}    &$\pm$  &\00.20$^e$		         &					   &	  	&		\\
 \\
 & $\Gamma$
            	 & 32\;\0\0\0&$\pm$  &\03$^{\dagger}$      	& \034.29$^{a}$  		   & 	   	& \032.22$^{b}$    \\
&          	&                  &             &                                        	&\027.81$^{g}$           	   &      	 & \035.20$^{h}$    \\
&          	&                 &              &                                         	&                                 	   	  &      	& \\
\\
 & $\overline{\sigma}^{\rm PI}$
             & 5.04\phantom{0} & $\pm$ & \01.0$^{\dagger}$   		& \0\04.50$^{a}$  		   &	 	&              \\
\\
 $\rm 1s2s^22p$ ~ $^1$P$^{\rm o}$					& $E_{\rm ph}^{\rm (res)}$
               	 & 194.484	&$\pm$  &\00.03$^{\dagger}$   & 194.485$^{a}$    		  & 	 	& 194.311$^{b}$  \\
&              	 & 194.39\phantom{0}	&$\pm$  &\00.10$^{d}$		& 194.670$^{i}$ 	           &	 	& 194.394$^{c}$ \\
&              	 & 194.532	&$\pm$  &\00.20$^{e}$		&					  &	 	& 194.493$^{f}$  \\
&                 & 194.36\phantom{0}        &$\pm$  &\00.10$^{k}$                &                                           &                &  194.543$^{j}$\\
 \\
 & $\Gamma$
             & 47\;\0\0\0&$\pm$  &\02$^{\dagger}$      		& \047.02$^{a}$  		  & 		& \045.40$^{b}$\\
 &         &		      &		       &					& \042.42$^{g}$	  	  &      	& \048.70$^{h}$\\
 &          &                 &                 &                                         	&                   		  	  &     	& \055.50$^{j}$\\			
\\
 & $\overline{\sigma}^{\rm PI}$
             & 42.1\0\0 & $\pm$ & \08.5$^{\dagger}$   		& \025.09$^{a}$  		  &	 	&              \\
\br
\end{tabular}
~\\
$^{\dagger}$Present ALS experimental results.\\
$^{a}$R-matrix with pseudo-states (RMPS) $LS$-coupling;  present work.\\
$^{b}$Saddle-Point-Method with Complex Rotation (SPM+CR) \cite{Lin2001,Lin2002}.\\
$^{c}$Dirac-Fock single configuration calculations \cite{Costello1992}.\\
$^{d}$Laser-produced plasma experiment \cite{Costello1992}.\\
$^e$Electron spectroscopy experimental data \cite{rodbro79} revised \cite{chung83,chung90};
the error bars of 0.2~eV are the estimates from the work of Chung and Bruch \cite{chung83}
who provide the Auger energies with 5-digit numbers.\\
$^f$Saddle-Point-Method (SPM) \cite{chung83,chung89}.\\
$^{g}$R-matrix  $LS$-coupling  \cite{Petrini1981}.\\
$^h$1/Z Perturbation \cite{Safron80}.\\
$^i$R-matrix  $LS$ -coupling \cite{Berrington1997}.\\
$^j$Complex scaling, multi-reference configuration interaction (MR-CI) \cite{Yeager2012}.\\
$^{k}$Inferred from existing experiments employing theoretical modeling \cite{Ryabtsev2005}.\\
\end{indented}
\end{table}

Experimental and theoretical resonance strengths $\overline{\sigma}^{\rm PI}$
were obtained (in the appropriate photon energy ranges) for the B$^+$ ion by integrating the cross section
 $\sigma_{\rm PI} ({\rm E}, i \rightarrow j )$ for a specific resonance transition from state $i$ to state $j$ over all energies
\begin{equation}
\overline{\sigma}^{\rm PI} = \int_{E}{}  \sigma_{\rm PI} ({\rm E}, i \rightarrow j )d{\rm E}\quad.
\end{equation}
The Voigt profiles fitted to the experimental results in figure~\ref{fig:Fig22meV}c) yield not only the
natural linewidths of the three prominent resonances  found
 in the photon energy range 193.60--194.7 eV but also allow for the extraction of further
 resonance parameters, i.e., energies $E_{ph}^{(res)}$ and strengths $\overline{\sigma}^{\rm PI}$.
Results for these 3 resonances are presented in table~1 along with data from previous studies.
In general, there is quite satisfying agreement between the different data sets.
As far as the resonance energies are concerned, the present R-matrix results are generally
closer to the present experiment than previous calculations, when available.
The present experimental resonance energies for the 3 terms addressed in table~1 have
at most 30 meV uncertainty which is by factors of 3 to 7 lower than the uncertainties of
all experimental data published previously. The differences of $E_{ph}^{(res)}$ between
the present theory and experiment are 1~meV for the $^3$P, 6~meV for the $^3$D and
1~meV for the $^1$P$^o$ resonances. While this is within the absolute experimental
error of $\pm 30$~meV, the relative experimental uncertainty of the peak positions with respect to one
another is only about 1~meV. The experimental energy difference between the $^3$P and
the $^3$D resonance is $225 \pm 2$~meV compared to $228$~meV obtained by the
present R-matrix with pseudostates (RMPS) calculations and the experimental energy difference between
the $^3$P and the $^1$P$^{\rm o}$ resonance amounts to $641 \pm 2$~meV which
 the present theoretical calculations perfectly reproduce.
In contrast to this the Saddle-Point-Method with Complex Rotation yields
the numbers 252~meV and 438~meV, respectively,
which are significantly further off than the present calculations.

For the natural linewidths of the resonances shown in table~1 there is excellent
agreement of the experimental and theoretical results of the present work for the
$^3$P, $^3$D and the $^1$P$^{\rm o}$ resonances. This is also true for most of the other theoretical
data. Only two calculations are significantly outside the experimental error bars. The complex scaling calculation
by Zhang and Yeager~\cite{Yeager2012} using multi-reference configuration interaction (MR-CI) yielded a
natural width for the $^1$P$^{\rm o}$ level 8.5~meV above the experiment which is more than 4.2 times
the experimental error bar. For the $^3$D resonance the complex scaling MR-CI method predicting
a width almost 4.2~meV  below the experiment is 40 \% outside the experimental error bar.
The R-matrix $LS$-coupling calculation of Petrini~\cite{Petrini1981} on the the $^1$P$^{\rm o}$ level
resulted in a natural width 4.6~meV below the experiment which is 2.3 times the experimental error bar.
In the case of the $^3$D resonance, the theoretical predictions differ from one another by up to 20\% but only up to
about 8\% from the present experiment which has a relative uncertainty of slightly less than 10\%.

PI resonance strengths are only available from the present work. With the assumption of a 40\% metastable
fraction present in the experiment, chosen to provide an optimum overall match of the present
theoretical and experimental data, the resonance strengths from the present R-matrix calculation
are at most 10\% off the experimental findings for the $^3$P and the $^3$D resonances.
Considering the estimated total experimental uncertainty of the absolute cross-section
measurement this means perfect agreement of theory and experiment.
For the $\rm 1s2s^22p$~$^1$P$^{\rm o}$ resonance, however, we note that the
theoretical value for the resonance strength is considerably lower (by almost 40\%) than
experiment. At this point one might question the assumption of a 40\% metastable
fraction in the ion beam. Scepticism might be supported
by the fact that Schippers \etal ~\cite{Schippers2003} assumed
a metastable fraction of 29\% in the ion beam used to measure PI for the
valence shell. Indeed, the resonance strength of the $^1$P$^{\rm o}$
resonance populated from the ground state of B$^+$  would  go up with the
assumption of a 29\% metastable fraction by almost 20\%, however, the good agreement
found for the other two resonances would be spoiled. Moreover, it was discussed recently
on the basis of detailed studies on Be-like ions~\cite{Mueller2010b} that those previous
findings were influenced by missing oscillator strengths in the accompanying calculations
due to an insufficient resolution of the resonances in the calculations. Moreover, the comparison of the
present PI results for the 1s2s2p$^2$~$^3$D resonance with the time-reversed
recombination of B$^{2+}$ into the identical intermediate state (see below) provides
strong evidence for the 40\% fraction being correct. Since the three resonances were
measured in several energy scans covering the whole energy range
from 193.7--194.7~eV their relative heights are very well determined
by the experiment, the partial discrepancy between the present theory and
the experimental data cannot be removed by assuming different values for
the metastable fraction of ions in the parent beam.

\begin{figure}
\begin{center}
\includegraphics[width=0.6\textwidth]{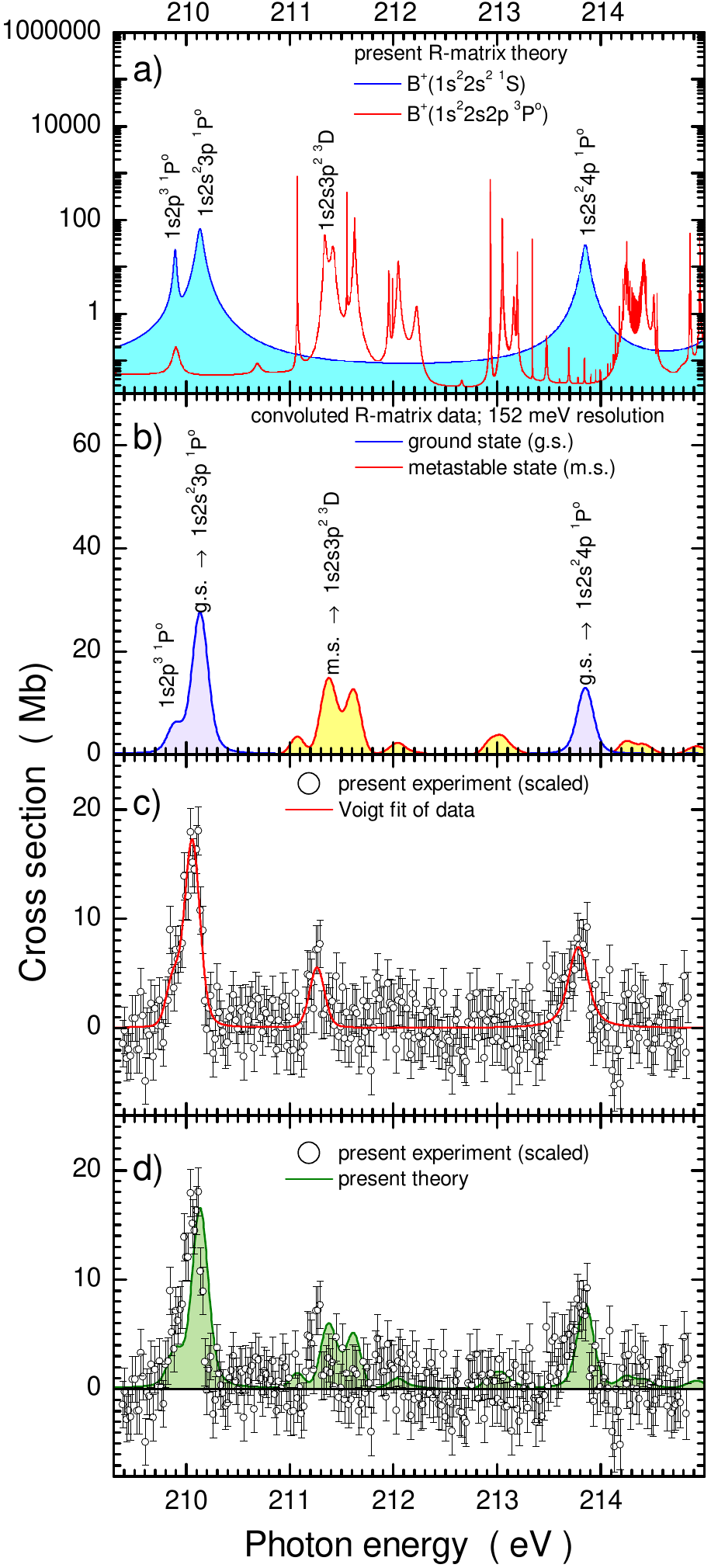}
\caption{\label{fig:Fig152meV} Experimental and theoretical cross sections for {\it K}-shell
					photoionization (PI) of Be-like B$^\mathrm{+}$ ions.
                                            The present R-matrix results a) for  ground-state (dark solid line with shading) and  metastable-state contributions
                                            (light solid line) were convoluted with a 152~meV FWHM Gaussian distribution function  to simulate the ALS
                                            experimental resolution and are then shown on a linear scale in b).  Designations of the strongest resonances
                                            populated from the ground state (g.s.) and the metastable state (m.s.) are provided in a) and b).
                                            The open circles in c) are the experimental results, the solid line is a 4-peaks Voigt fit to the data.
                                            A comparison of the experimental data and the R-matrix results
                                            weighted as 60\% ground and 40\% metastable states is provided in d).}
\end{center}
\end{figure}

Figure~\ref{fig:Fig152meV} addresses a second energy range (from about 209--215~eV) that was
investigated in the present experiments. In this energy region an overview scan of the PI cross section
was carried out at 152~meV resolution. Hence, the natural cross sections calculated
for the B$^{+}$(1s$^2$2s$^2$~$^1$S) ground-state and the B$^{+}$(1s$^2$2s2p~$^3$P$^{\rm o}$)
metastable-state components of the parent ion beam (see panel a) ) were convoluted with a
152 meV FWHM Gaussian and displayed in panel b). On the basis of this R-matrix result only
three prominent peaks are to be expected in an experimental spectrum covering this energy range.
Indeed, the experimental spectrum exhibits 3 peak features. However, the peak at about 210~eV is
sufficiently asymmetric to call for the presence of an additional, in this case unresolved, fourth resonance.
Accordingly, a 4-peaks Voigt profile distribution  was employed to fit the experimental data
displayed in panel c). The experimental data were determined on a relative scale.
For display in figure~\ref{fig:Fig152meV}c) and ~\ref{fig:Fig152meV}d) they were normalized
to the total theoretical resonance strength predicted for the energy range of 209--210.6~eV.
The bottom panel \ref{fig:Fig152meV}d) shows a comparison of the normalized experimental
data with the appropriately convoluted R-matrix calculations assuming a 40\% fraction
of $^3$P$^{\rm o}$ metastable B$^+$ ions admixed to 60\% of $^1$S ground-state ions
in the parent ion beam. In spite of the relatively modest resolution of 152~meV, differences
between the measured spectra and the R-matrix calculations can be seen.
Over all, the present calculations are in suitable agreement with the experiment.

\begin{figure}
\begin{center}
\includegraphics[width=0.6\textwidth]{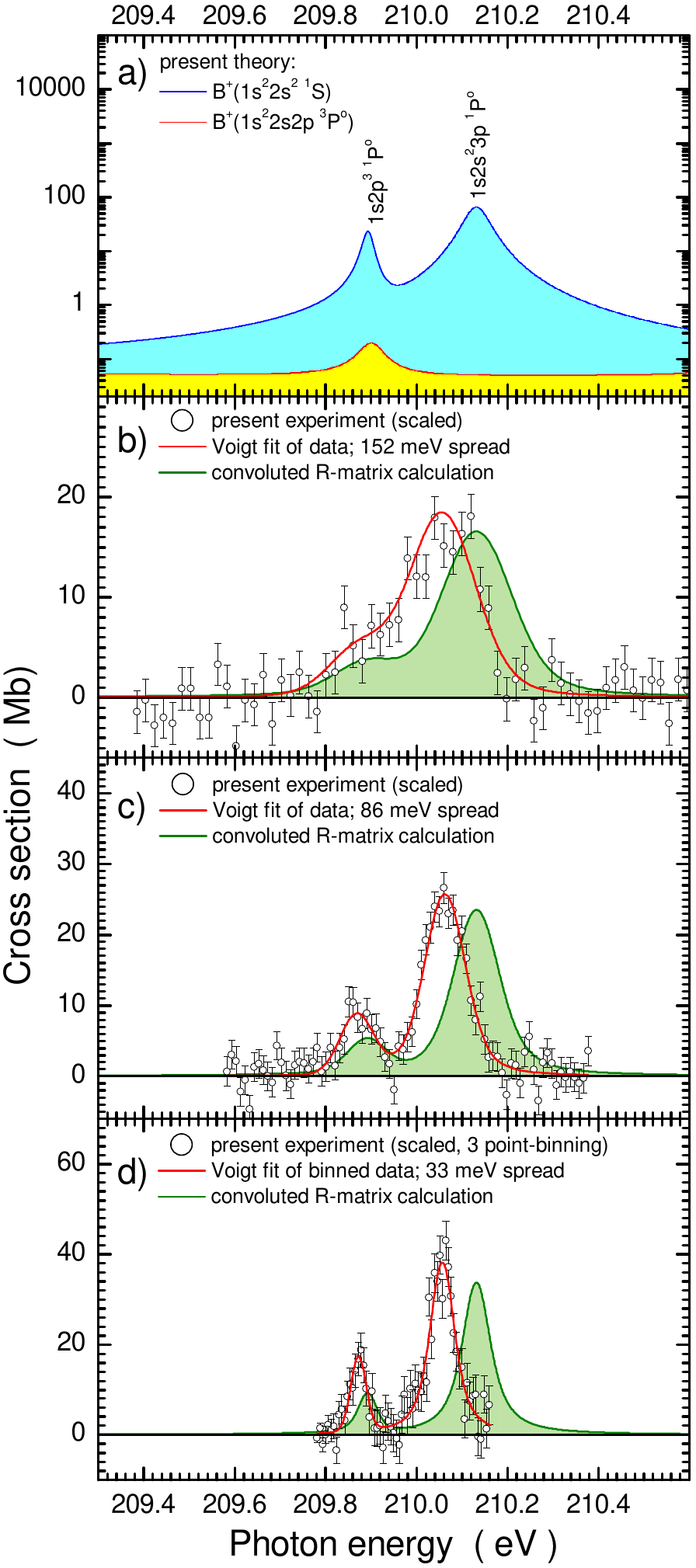}
\caption{\label{fig:3resolutions} Experimental and theoretical cross sections for {\it K}-shell
                        photoionization (PI) of Be-like B$^\mathrm{+}$ ions. The present R-matrix results for ground-state
                        (dark solid line with shading) and metastable-state contributions (light solid line with light shading) are shown in a).
                        Panels b), c) and d) provide measurements (open circles) at resolutions 152~meV, 86~meV and
		     33~meV, respectively, together with Voigt fits (solid line through the
		     experimental data) and appropriately convoluted R-matrix results.
		     The R-matrix data in b), c) and d) were weighted as 60\% ground and
                        40\% metastable states to simulate the experimental results.}
\end{center}
\end{figure}

The effect of energy resolution on the resonance features in the energy range 209.3--210.6~eV
is illustrated in figure~\ref{fig:3resolutions}. Panel a) shows the natural
(unconvoluted) cross sections resulting from the present R-matrix calculations.
The calculations were carried out for the parent B$^+$ ion in its ground state or in the $^3$P$^{\rm o}$ metastable state.
The comparison of theory and experiment  at different levels depending on the experimental resolution are shown in panels b)
for 152~meV resolution, c) for 86~meV resolution, and d) for 33~meV resolution.
%
%
%

\begin{table}
\caption{\label{tab:fit2} Resonance energies $E_{\rm ph}^{\rm (res)}$ (in eV),
         and  natural linewidths $\Gamma$ ( in meV) for B$^{+}$ ions in the photon energy region
         209--215 eV from the present  experimental and theoretical investigations compared
         with previous studies. Although the present experimental uncertainty of energies on an
         absolute scale is $\pm 30$~meV,  three decimal figures are provided for the resonance
         energies  because the experiment is sensitive to the 1~meV level and allows one to
         determine energy spacings with uncertainties of the order of 2~meV. }
\begin{indented}
 \lineup
 \item[]\begin{tabular}{ccr@{\,}c@{\,}llcl}
\br
 Resonance    & & \multicolumn{3}{c}{ALS/Others}   & \multicolumn{1}{c}{R-matrix} & \multicolumn{2}{c}{Others}\\
 (Label)      & & \multicolumn{3}{c}{(Experiment)}      & \multicolumn{1}{c}{(Theory)} & \multicolumn{2}{c}{(Theory)}\\
 \ns
 \mr
 \\
$\rm 1s2p^3$ ~ $^1$P$^{\rm o}$	 					 & $E_{\rm ph}^{\rm (res)}$
               & 209.872   &$\pm$  & \00.03$^{\dagger}$  		& 209.895$^{a}$  		& 	& 209.894$^{b}$			 \\
&            & 		&  		&   						& 210.053$^{e}$		& 	&    \\
\\
 & $\Gamma$
            & 4 \0\0\0  	& $\pm$  & \014$^{\dagger}$			  & \026$^{a}$		& 	&\027$^{b}$ \\
 &              &    	&     &  		     					  & \021$^{e}$		& 	&   \\
    \\
  \\
 $\rm 1s2s^23p$ ~ $^1$P$^{\rm o}$					 & $E_{\rm ph}^{\rm (res)}$
              & 210.057	&$\pm$  &\00.03$^{\dagger}$   	& 210.130$^{a}$    		& 	& 210.125$^{c}$  \\
&           & 210.14\phantom{0}	&$\pm$  &\00.10$^{d}$			& 209.843$^{e}$		&	&  \\
&           & 210.11\phantom{0}	&$\pm$  &\00.10$^{f}$			& 		&	&  \\
\\
 & $\Gamma$
             & 43\;\0\0\0& $\pm$  &\09$^{\dagger}$     			& \059  		 & 	& \0 --   \\
 \\
\\
 ($\rm 1s2s3p^2$ ~ $^3$D)							 & $E_{\rm ph}^{\rm (res)}$
              & 211.259   &$\pm$  & \00.03$^{\ddagger}$  		& 211.369$^{a}$  		& 	&\0 -- \\
\\
 & $\Gamma$
            &    \;\0\0\0 & --  &     		& \012$^{a}$	& 	&\0 --\\
   \\
   \\
 $\rm 1s2s^24p$ ~ $^1$P$^{\rm o}$					 & $E_{\rm ph}^{\rm (res)}$
              & 213.784	&$\pm$  &\00.03$^{\ddagger}$   	&213.853$^{a}$    		& 	&213.715$^{c}$ \\
&           & 213.76\phantom{0}	&$\pm$  &\00.10$^{d}$			&213.680$^{e}$		&	& \\
&           & 213.73\phantom{0}	&$\pm$  &\00.10$^{f}$			&					&	& \\
 \\
 & $\Gamma$
             & \;\0\0\0 & --		&\0     	& \064$^{a}$  		 & 	& \0 --   \\
              \\
\br
\end{tabular}
~\\
$^{\dagger}$Present ALS experimental results taken at 33 meV.\\
$^{\ddagger}$Present ALS experimental results taken at 152 meV.\\
$^{a}$R-matrix  with pseudo-states (RMPS) $LS$-coupling  present work.\\
$^{b}$Saddle-Point-Method with Complex Rotation (SPM+CR) \cite{Lin2001,Lin2002}\\
$^{c}$Dirac-Fock single configuration calculations \cite{Costello1992}\\
$^{d}$Laser-produced plasmas experiment \cite{Costello1992}\\
$^{e}$R-matrix  $LS$-coupling  \cite{Petrini1981}.\\
$^{f}$Inferred from existing experiments employing theoretical modeling \cite{Ryabtsev2005}.\\
\end{indented}
\end{table}

From the Voigt fits shown in figures~\ref{fig:Fig152meV} and \ref{fig:3resolutions} resonance parameters
were determined as far as possible.  They are compared in table~2 with the R-matrix results and with other methods.
With the limited statistical quality of the experimental data and the missing absolute calibration,
the comparison between theory and experiment is not as exhaustive as for the stronger resonances addressed in table~1.
The agreement between theory and experiment for the more highly excited intermediate
states is not as good as that for the lower states. We note also that comparisons between different theoretical
approaches for resonance energies may be made and for a single case theoretical
natural widths can be compared. Results from the present calculation and the
Saddle-Point-Method with Complex Rotation are in comparable agreement for
the 1s2p$^3$~$^1$P$^{\rm o}$ resonance.
For more definitive results, better experiments would be necessary.

\begin{figure}
\begin{center}
\includegraphics[width=0.6\textwidth]{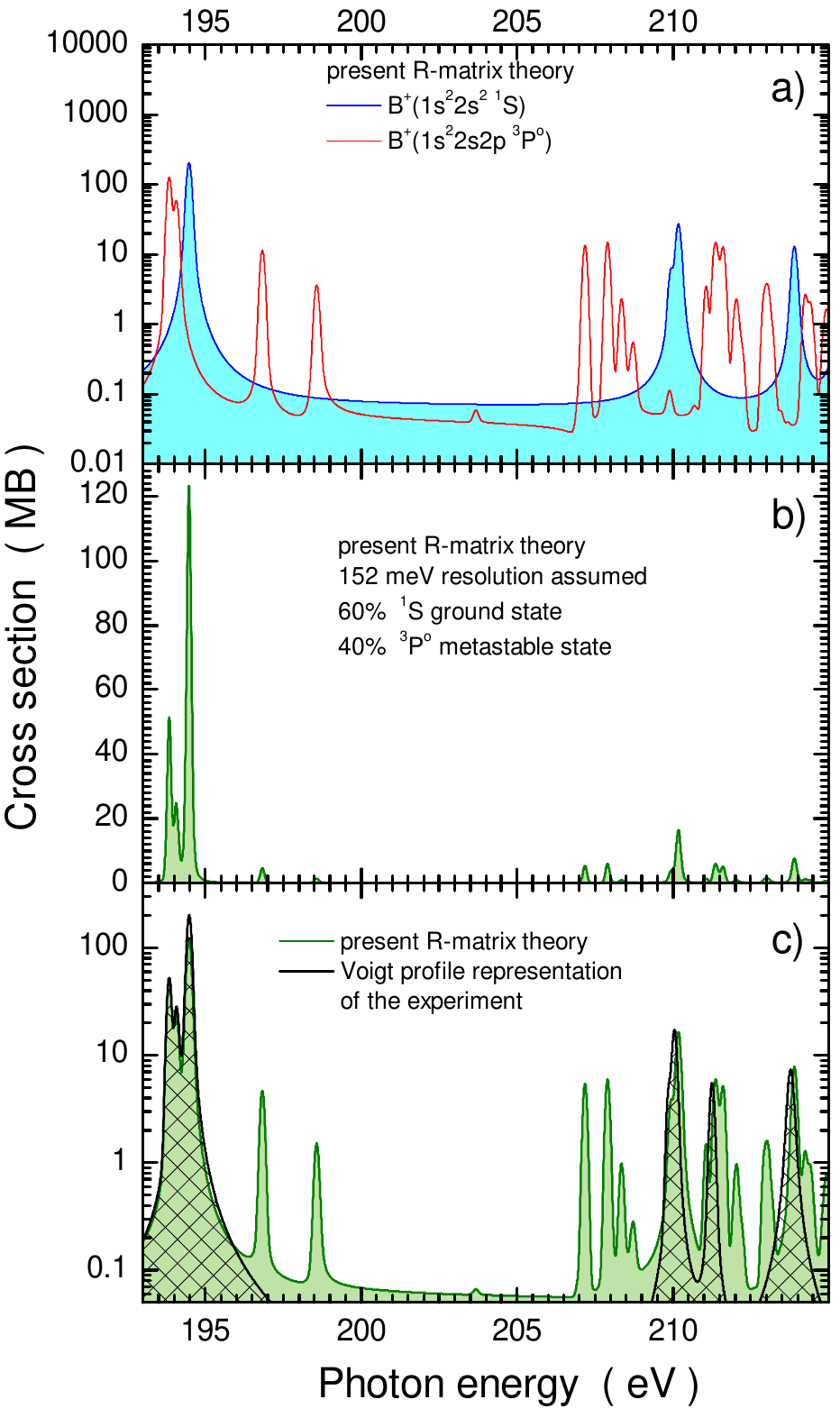}
\caption{\label{fig:152meVsummary}  Overview of the cross sections for
                                            {\it K}-shell photoionization (PI) of Be-like B$^\mathrm{+}$ ions that could
                                            be experimentally accessed in the present experiments. Panel a)
                                            shows the present R-matrix results for ground-state
                                            (dark solid line with shading) and metastable-state contributions (light solid line)
                                            convoluted with a Gaussian 152~meV FWHM distribution function.
                                            Panel b) displays the simulation of an experiment at 152~meV
                                            resolution assuming 60\% ground and 40\% metastable states
                                            in the parent ion beam. Panel c) makes obvious that the 7 strongest
                                            peaks were experimentally observed. With the resonance parameters
                                            available from the Voigt fits of experimental data in the previous figures
                                            an overview experiment at 152~eV is simulated (black solid line with
                                            cross-hatched shading) and compared in c) with the theory data
                                            from panel b). Note that panel b) has a linear scale to show the
                                            relative sizes of the PI resonances, while a) and c) are
                                            on log scales illustrating the presence of numerous resonances.}
\end{center}
\end{figure}

Figure~\ref{fig:152meVsummary} provides an overview and a summary of the resonance
structures found in the experiments and permits an overall comparison of the present
experimental and theoretical results at an assumed constant energy resolution of 152~meV.
Panel a) displays the R-matrix data for ground-state and metastable-state
parent ions convoluted with a 152~meV FWHM Gaussian distribution function.
For showing the relative strengths of the resonance features in the energy
range 194--215~eV panel b) displays the weighted sum of the
R-matrix data for the ground state (60\%) and the metastable state (40\%).
Clearly, the 3 resonances at around 195~eV dominate the spectrum.
The resonances in the range 209--215~eV are roughly an order
of magnitude smaller. The experiment detected just the strongest features
in that range and missed smaller resonances near 197~eV and 208~eV.
However, as panel c) illustrates, the experiment could determine energies
and resonance strengths (though only relative for the smaller resonances)
that show very satisfying agreement with the theoretical predictions at the
level of detail that can be accessed in this overview.

Cross-section calculations were also carried out for the ground-state photoionization of this B$^{+}$ ion,
with a larger number of states (390-levels) retained in the close-coupling calculations with the current n=4 basis set.
These calculations were undertaken to address the discrepancy in the strength of the $^1$P$_1$ resonance
located at  approximately 194.5 eV. The 390-level calculations gives  results very similar to those from using  249-levels and the two
data sets cannot be distinguished from one another when convoluted at the 22 meV FWHM experimental resolution.
Due to the limited size of the basis set used in the R-matrix $LS$-coupling calculations, we speculate that one
would require a much larger basis set, or that cross section calculations may need to be performed in intermediate coupling
to bring the peak strength of this resonance into closer agreement with experiment. A similar analogy can be
drawn with the recent {\it K}-shell study on neutral atomic oxygen \cite{Stolte2013}.

We point out for present day astronomical x-ray observations, the instruments
carried by satellites such as {\it XMM-Newton} and {\it Chandra} have energy resolutions of at best $\sim$ 0.6 eV (600 meV).
Radiation facilities such as the ALS, SOLEIL, ASTRID II, BESSY II or PETRA III,
provide much higher resolution and greater precision capabilities.
Furthermore, there are various issues concerning the calibration of spectra obtained from satellites.
Considering this, the predictive strength of the R-matrix calculations is
well suited for analyzing measured  spectra of astrophysical objects when benchmarked
against high resolution experimental measurements such as in the present investigation.

\subsection{Time-reversal symmetry}
The present results obtained for photoionization of B$^+$ ions can partly be
 related to an electron-ion recombination experiment~\cite{Boehm2006,Lestinsky2004}
 carried out with a cooled beam of B$^{2+}$(1s$^2$2s~$^2$S) ions at the heavy-ion
 storage-ring TSR of the Max-Planck-Institute for Nuclear Physics in Heidelberg,
 Germany.

 Time reversal symmetry and the principle of detailed
 balance/micro-reversibility~\cite{Flannery2006}
 relate cross sections $\sigma^{\mathrm{PI}}$ for photoionization
\begin{equation}
 \gamma + |i\,\rangle \rightarrow  |f\,\rangle + e - I_{\mathrm{bind}}
		\label{eq:PI}
\end{equation}
and cross sections $\sigma^{\mathrm{PR}}$ for (photo)recombination
\begin{equation}
 		 e + |f\,\rangle \rightarrow  |i\,\rangle + \gamma + I_{\mathrm{bind}}
		\label{eq:PR}
\end{equation}
on a level-to-level basis ($|i\,\rangle \rightarrow  |f\,\rangle$ and $|f\,\rangle \rightarrow  |i\,\rangle$
with well defined levels $|i\,\rangle$ and $|f\,\rangle$). Here, $\gamma$ denotes a photon with
energy $E_{\mathrm ph}$. The binding energy $I_{\mathrm{bind}}$ is the difference of the
total energies of the final and the initial levels $I_{\mathrm{bind}}=E(|f\,\rangle)-E(|i\,\rangle)$.
The energy of the photoelectron (in the electron-ion center-of-mass system) is
$E_{\mathrm{e}} = E_{\mathrm ph} - I_{\mathrm{bind}}$.
The principle of detailed balance~\cite{Flannery2006} yields the following relation for nonrelativistic
photon energies $h\nu \ll m_{\textrm{e}}c^2$
\begin{equation}\label{eq:balance}
   \frac{ \sigma^{\mathrm{PR}}_{f \to i}}{\sigma^{\mathrm{PI}}_{i \to f}} =
     \frac{g_i}{g_f}\frac{E_\mathrm{ph}^2}{2m_{\mathrm{e}}c^2E_{\mathrm{e}}}
\end{equation}
where the quantities $g_i$ and $g_f$ are the statistical weights of the initial and final levels,
respectively, $m_{\mathrm{e}}$ is the electron rest mass and $c$ the vacuum speed of light.

Exploiting time-reversal symmetry in photoionization and photo-recombination is not generally
straight forward~\cite{Mueller2008,Mueller2002,Schippers2002,Schippers2004,Mueller2009}. Usually in the photoionization experiments with ions, the final level $|f\,\rangle$
is not specified and even the initial level $|i\,\rangle$ may have some ambiguity as is the case
in the present study with ground state and metastable initial ions. While photo-processes for
light atoms and ions obey relatively strict selection rules, recombination permits many more
individual reaction pathways. Electron-ion recombination via resonances produces intermediate
multiply excited states. These states mostly stabilize by the sequential emission of two photons.
The time-reversed process for that would be a two-photon ionization process. Clearly, this is
out of the reach of most (all) photoionization experiments (with ions). Considering these limitations it becomes
clear, that most of the features found in either photoionization or photo-recombination
experiments cannot be related to one another on the basis of detailed balancing
because the appropriate pathway was not observed in both directions.

In the present case, for photoionization of B$^+$ and  photo-recombination of B$^{2+}$ ions, only
one resonance can be found in the whole energy range investigated that allows for applying the
principle of detailed balance. This is the 1s2s2p$^2$~$^3$D resonance that can be populated
by photo-excitation from the metastable B$^+$(1s$^2$2s2p~$^3$P$^{\rm o}$) level and by
dielectronic capture being the first step of resonant B$^{2+}$(1s$^2$2s~$^2$S)~+~e$^{-}$ recombination.
In the present experiment a 40\% fraction ($f=0.4$) of the parent ion beam in the
B$^+$(1s$^2$2s2p~$^3$P$^{\rm o}$) term ($|i\,\rangle$) produces the B$^+$(1s2s2p$^2$~$^3$D)
resonance by photo-excitation at an energy $E_\mathrm{ph}~=~194.08~\pm 0.03$~eV. The excited
state decays by Auger processes to either B$^{2+}$(1s$^2$2p~$^2$P)~+~e$^{-}$, with a branching
factor $\omega_A^{2p}~=~0.109$~\cite{Lin2001}, or to B$^{2+}$(1s$^2$2s~$^2$S)~+~e$^{-}$, with a
branching factor $\omega_A^{2s}~=~0.891$~\cite{Lin2001}. In the storage-ring recombination
experiment~\cite{Boehm2006,Lestinsky2004} 100\% of the parent ions are in the
B$^{2+}$(1s$^2$2s~$^2$S) ground level ($|f\,\rangle$). By dielectronic capture of the
incident electron at energy $E_{\mathrm{e}}~=~173.26$~eV the B$^+$(1s2s2p$^2$~$^3$D)
resonance is formed. For the recombination process to be completed, this resonance
has to stabilize by photoemission. Intercombination decay to the B$^+$(1s$^2$2s$^2$~$^1$S)
ground state is forbidden by the electric dipole selection rules. Thus, stabilization inevitably
populates (with a probability of 100\%) the metastable B$^+$(1s$^2$2s2p~$^3$P$^{\rm o}$) level.
The resonance energy expected for the B$^{2+}$ recombination is
$E_{\mathrm{e}}~=~E_{\mathrm ph}~-~I_{\mathrm{bind}}~=~194.08~{\rm eV}~-~20.52~{\rm eV}~=~173.56~{\rm eV}$
just about 0.3~eV above the storage-ring experiment. In the present case $I_{\mathrm{bind}}$ is the
 binding energy of the 2p electron in the metastable
 B$^+$(1s$^2$2s2p~$^3$P$^{\rm o}$) ion~\cite{NIST2013}.
 The discrepancy of 0.3~eV is within the uncertainty of the experimental
 energy calibration of the storage-ring experiment.

\begin{figure}
\begin{center}
\includegraphics[width=0.8\textwidth]{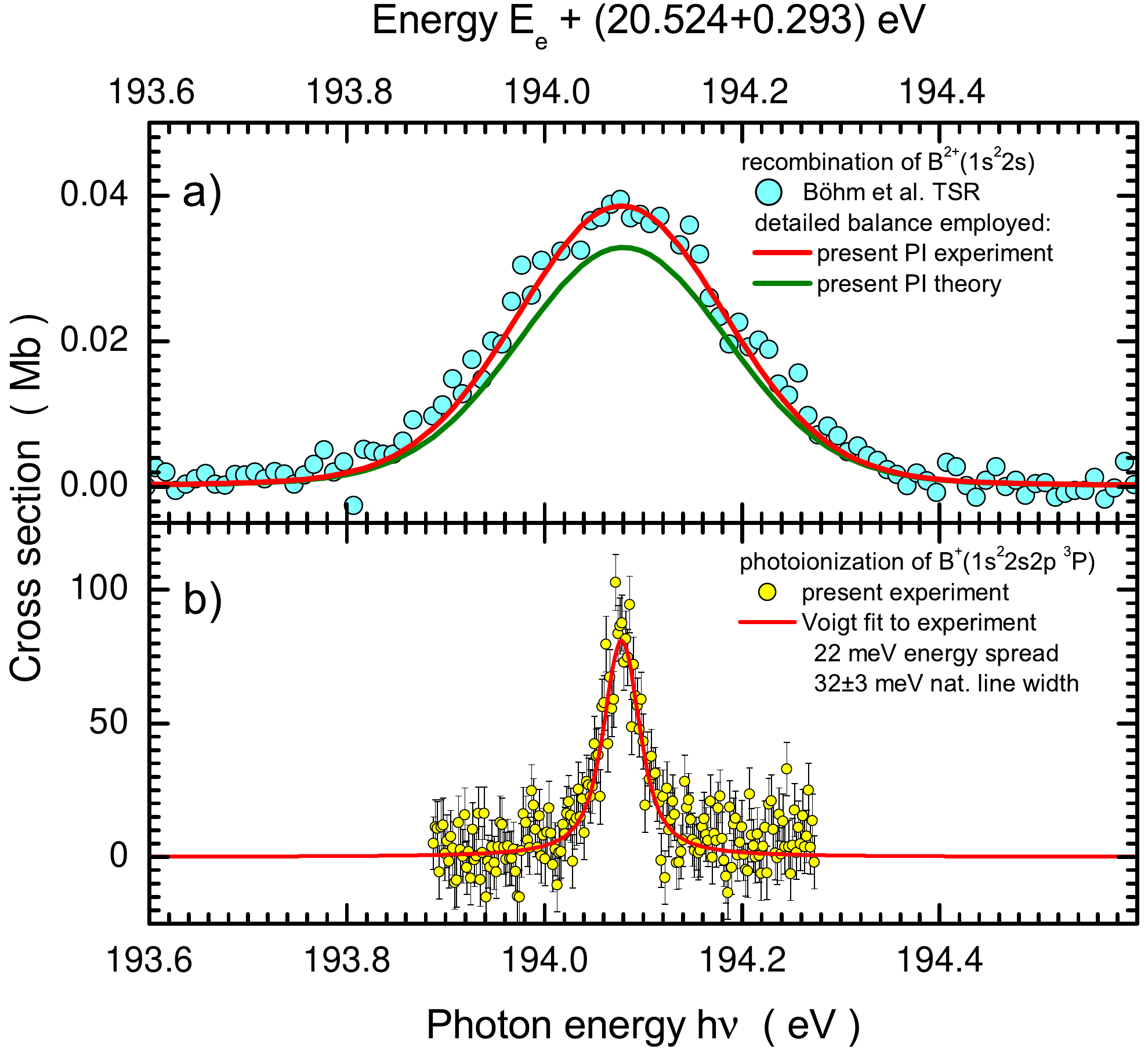}
\caption{\label{fig:detailedbalance}  Absolute cross sections for the population of the 1s2s2p$^2$~$^3$D
							resonance by a) recombination of B$^{2+}$(1s$^2$2s~$^2$S) with
							an electron \cite{Boehm2006,Lestinsky2004} and b) photo-excitation
							of the B$^{+}$(1s$^2$2s2p~$^3$P$^{\rm o}$) ion (present experiment).
							The measured data are represented by shaded circles.
							The solid line in b) is the Voigt fit to the PI cross section.
							The light solid curve in a) almost perfectly coinciding with the
							experimental data was obtained by applying detailed balance
							to the Voigt fit from panel b). The dark solid line results from applying
							detailed balance to the present R-matrix theory result
							assuming a 40\% metastable ion fraction in the PI experiment.}
\end{center}
\end{figure}

The lower panel b) of figure~\ref{fig:detailedbalance} shows the result of the present
photoionization experiment in the energy range where B$^+$(1s2s2p$^2$~$^3$D) is
excited from the metastable B$^+$(1s$^2$2s2p~$^3$P$^{\rm o}$) term. The solid line
is a Voigt fit representing the experimental cross section $\sigma^{\mathrm{PI}}_{ms}(E_\mathrm{ph})$.
The upper panel a) of figure~\ref{fig:detailedbalance} displays data from the
previous photo-recombination experiment at the Heidelberg storage ring~\cite{Boehm2006,Lestinsky2004}
using B$^{2+}$(1s$^2$2s~$^2$S) ground-level ions. The electron energy axis is shifted by the
ionization energy $I_{\mathrm{bind}}= 20.52$~eV to account for the binding energy of the
2p electron in the metastable B$^+$(1s$^2$2s2p~$^3$P$^{\rm o}$) ion and an additional
correction of about 0.3~eV (see above) is applied to line up the resonance measurements
shown in the two panels of figure~\ref{fig:detailedbalance}. The energy spread in the storage
ring experiment was found to be about 0.23~eV, more than 10 times that of the PI experiment.
Note that the maximum of the recombination cross section is about 40~kb while the maximum
of the PI cross section is almost 100~Mb. In order to derive the recombination cross section
on the basis of the present PI cross section
represented by $\sigma^{\mathrm{PI}}_{ms}(E_\mathrm{ph})$ the principle of
detailed balance (equation~\ref{eq:balance}) can be applied in the form~\cite{Flannery2006}
\begin{equation}
    \sigma^{\mathrm{PR}}_{f \to i} = \omega_A^{2s} \frac{\sigma^{\mathrm{PI}}_{ms}(E_\mathrm{ph})}{f}
     \frac{g_i}{g_f}\frac{E_\mathrm{ph}^2}{2m_{\mathrm{e}}c^2E_{\mathrm{e}}}.
\end{equation}
With the branching ratio $\omega_A^{2s} = 0.891$, the metastable fraction $f=0.4$, the statistical
weight $g_i=9$ of the initial (metastable) $^3$P$^{\rm o}$ term and the statistical weight $g_f=2$
of the final B$^{2+}$(1s$^2$2s~$^2$S) term the expected recombination cross section is obtained
at an energy resolution of 32~meV. When this result is transformed to the width of 232~meV
to account for the energy resolution of the recombination experiment, the light solid line
in panel a) results which is in almost perfect agreement with the heavy-ion storage
ring recombination data. Also the present R-matrix cross section has been subject
to the same procedure and yields the slightly lower dark solid line. The agreement
found for two totally independent absolute cross section measurements with completely
different experimental arrangements, along with the R-matrix predictions, provides high confidence in the absolute
measurements and strongly supports the assumption of a
metastable fraction of 40\% in the present PI experiments.

\section{Summary and Conclusions}
Photoionization of Be-like boron ions, B$^{+}$, has been investigated
using state-of-the-art experimental and theoretical methods.
High-resolution spectroscopy was performed with
E/$\Delta$E = 8800 ($\sim$ 22 meV) covering the
energy ranges 193.7--194.7 eV and 209--215 eV, where several strong
peaks in the cross sections were found in proximity to the {\it K}-edge.
 For the  peaks observed in the lower-energy range, very good agreement is found with
 respect to resonance energies and natural line widths between the present theoretical and
experimental results. A smaller theoretical strength compared to experiment for the strongest resonance,
that is accessible by photo-excitation of the ground state, we attribute to the limited basis
set and to the limitations of the $LS$-coupling scheme applied in the calculations.
Closer to the {\it K}-edge where the states involved are
more highly excited, the agreement between theory and experiment is slightly
less favourable. However, at the level of resolution of astronomical observations
in the present photon energy range the present theoretical results can
be applied with good confidence to the analysis of astrophysical spectra.
The strength of the present study is in its high experimental
resolving power coupled with theoretical predictions made using the R-matrix method.

The principle of detailed balance was used to compare the present PI cross-section
measurements with previous experimental cross sections for the time-inverse
photo-recombination (PR) process. The excellent agreement found by comparing
two totally independent absolute measurements, one at a synchrotron light source,
the other at a heavy-ion storage ring, coupled with the R-matrix predictions, provides high confidence in the
accuracy of both the experimental and theoretical results.

Given that the present results have been benchmarked with high-resolution
experimental data and with various other theoretical and experimental methods
they would be suitable to be included in astrophysical modelling
 codes such as CLOUDY  \cite{Ferland1998,Ferland2003},
 XSTAR \cite{Kallman2001} and AtomDB \cite{Foster2012}
 that are used to numerically simulate the thermal and
 ionization structure of ionized astrophysical nebulae.

\ack
We acknowledge support by Deutsche Forschungsgemeinschaft under project number Mu 1068/10  and through
NATO Collaborative Linkage grant 976362 as well as by the US Department of Energy (DOE)
under contract DE-AC03-76SF-00098 and grant  DE-FG02-03ER15424.
C Cisneros acknowledges support from PAPIT-UNAM IN107912-IN102613, Mexico.
B M McLaughlin acknowledges support by the US
National Science Foundation through a grant to ITAMP
at the Harvard-Smithsonian Center for Astrophysics,
a visiting research fellowship from Queen's University Belfast
and the hospitality of  A. M. and S. S. during a recent visit to Giessen.
We thank John C Raymond and Randall K Smith
from the Harvard Smithsonian Center for Astrophysics for helpful discussions
on the astrophysical applications.
The computational work was carried out at the National Energy Research Scientific
Computing Center in Oakland, CA, USA, the Kraken XT5 facility at the National Institute
for Computational Science (NICS) in Knoxville, TN, USA
and at the High Performance Computing Center Stuttgart (HLRS) of the University of Stuttgart, Stuttgart, Germany.
We thank Stefan Andersson from Cray Research for his assistance and
advice with the implementation and optimization of the parallel R-matrix codes on the Cray-XE6 at HLRS.
The Kraken XT5 facility is a resource of the Extreme Science and Engineering Discovery Environment (XSEDE),
which is supported by National Science Foundation grant number OCI-1053575.
The Advanced Light Source is supported by the Director,
Office of Science, Office of Basic Energy Sciences,
of the US Department of Energy under Contract No. DE-AC02-05CH11231.
%
%
%
%
%
\section*{References}
\bibliographystyle{iopart-num}
\bibliography{bplus}
\end{document}